\documentstyle[aps,pre,epsf,multicol]{revtex}
\begin{document}
\draft
\title{Reconstructed Rough Growing Interfaces; \\
  Ridgeline Trapping of Domain Walls}

\author{ Chen-Shan Chin and Marcel den Nijs}

\address{ Department of Physics, University of Washington, P.O. Box 351560, \\
  Seattle, Washington 98195-1560}

\date{\today}

\maketitle

\begin{abstract}
  We investigate whether surface reconstruction order exists in
  stationary growing states, at all length scales or only below a
  crossover length, $l_{\rm rec}$.  The later would be similar to
  surface roughness in growing crystal surfaces; below the equilibrium
  roughening temperature they evolve in a layer-by-layer mode within a
  crossover length scale $l_{\rm R}$, but are always rough at large
  length scales.  We investigate this issue in the context of KPZ type
  dynamics and a checker board type reconstruction, using the
  restricted solid-on-solid model with negative mono-atomic step
  energies.  This is a topology where surface reconstruction order is
  compatible with surface roughness and where a so-called
  reconstructed rough phase exists in equilibrium.  We find that
  during growth, reconstruction order is absent in the thermodynamic
  limit, but exists below a crossover length $l_{\rm rec}>l_{\rm R}$,
  and that this local order fluctuates critically.  Domain walls
  become trapped at the ridge lines of the rough surface, and thus the
  reconstruction order fluctuations are slaved to the KPZ dynamics.
\end{abstract}

\begin{multicols}{2}
  \narrowtext

\section{introduction}

Equilibrium surface phase transitions have been a topic of research
for several decades.  Various types of critical behaviors are well
established both in theoretical models and actual experiments.  This
includes surface roughening~\cite{vBN,Wortis}, surface
melting~\cite{Dash,Dietrich}, and surface reconstruction
~\cite{King,VV,MdN-Pt}.  Moreover, the competition between these
phenomena leads to additional phases and phase transitions, like
disordered flat phases, preroughening transitions, and reconstructed
rough phases~\cite{MdN-PR1,MdN-PR2,MdN-King}.  Roughening induced
deconstruction in Pt(110) \cite{MdN-Pt,RVK} and preroughening induced
deconstruction in Si(110) type geometries are other examples of this
competition~\cite{MdN-Si}.

The theory of dynamic non-equilibrium processes like surface growth
has flourished during the last decade as well.  Several new types of
dynamic universality classes have been identified.  KPZ type growth is
one example~\cite{KPZ,HHZh,RSOS,KrSp,KANK,KK,CCS-MdN,Lassig}.
Unfortunately, in this area a gap seems to widen between theoretical
and experimental interests.  Theoretical oriented research tends to
focus on universal aspects of these processes, such as the large scale
properties of growing surfaces in the stationary growing state and on
how this state is approached in the asymptotic large time limit.
Experimental oriented research tends to focus on more microscopic
short distance aspects of growing surfaces, e.g., as encountered in
actual epitaxial growth.

One of the fundamental issues, relevant to both perspectives, is
whether any of the above equilibrium surface phase transitions persist
in the stationary state of growing interfaces.  In this paper we
address whether surface reconstruction order can exist during growth.

This issue is related to the absence of surface roughening transitions
in growing surfaces. Below the equilibrium roughening transition
temperature $T_{\rm R}$ the growing surface is rough at large length
scales, but remains flat and grows layer-by-layer at distances shorter
than a crossover length scale $l_{\rm R}$, which varies with
temperature and oversaturation.  We review this briefly in
section~\ref{nucleation} in the context of elementary nucleation
theory.

Consider a surface that is flat and reconstructed in equilibrium at
low temperatures.  Below $T_{\rm R}$ it appears to grow within $l_{\rm
  R}$ as flat in a layer-by-layer mode.  Moreover, below $T_{\rm rec}$
(if $T_{\rm rec}<T_{\rm R}$) it appears as reconstructed if the new
particles can find their proper reconstruction positions at times
scales that are short compared to the rate at which a new layer is
completed. Presume that this is indeed the case.  The next, more
intriguing question is whether $l_{\rm rec}$ can be larger than
$l_{\rm R}$; i.e., whether rough growing surfaces be reconstructed?
The compatibility of surface roughness with surface reconstruction was
addressed in the context of equilibrium phase transitions several
years ago.  The answer depends on intricate details of the surface
topology.  For example, in missing row reconstructed (MRR) (110)
facets in FCC crystals, like Au and Pt, roughness is incompatible with
reconstruction order, and the surface roughening transition must
destroy the reconstruction simultaneously~\cite{MdN-Pt}.  In such
geometries, the reconstruction order can not exist in growing surfaces
beyond the roughness length scale either, and $l_{\rm rec}\leq l_{\rm
  R}$.

Surface roughness and reconstruction are compatible with each other in
other crystal structures.  Simple cubic (SC) MR reconstructed (110)
facets are an example.  In equilibrium, they can roughen before the
reconstruction order deconstructs, $T_{\rm R}<T_{\rm rec}$.  The
intermediate phase is known as a reconstructed rough
phase~\cite{MdN-Pt,MdN-King}.  For those surfaces it might be possible
to observe genuine deconstruction type phase transitions in growing
surfaces. Or, if not, the surface reconstruction can at least persist
well beyond the roughness crossover length scale, $l_{\rm rec}>l_{\rm
  R}$, and will be limited by an independent mechanism.  These issues
are the topic of our research reported here.

In section~\ref{nucleation} we review rough versus layer-by-layer
growth in surfaces, and in section~\ref{recrough} the basic properties
of equilibrium reconstructed rough phases.  Next, in section
\ref{recroughgrowth}, we start to focus on the reconstruction versus
dynamic roughness issue, and then, in section~\ref{RSOSmodel}, we
choose a specific type of reconstruction and a specific type of
surface growth dynamics to study it quantitatively by means of Monte
Carlo (MC) simulations.  The model must be as simple as possible,
avoiding secondary effects that might obscure the central issue.  Our
choice is the so-called restricted solid-on-solid (RSOS) model with
negative step energies, which describes a simple cubic checker board
type reconstruction, and KPZ type growth.  The MC simulation results
are presented in section~\ref{MCdata}, and analyzed in
section~\ref{trappedloops}.  Finally, in section~\ref{conclude} we
summarize our results.

\section{Roughness in growing surfaces}\label{nucleation}

The topic of this paper is whether surface reconstruction order can
exist during growth, but as a start it is useful to review briefly the
related issue of dynamic surface roughness from a long and short
length scale perspective.  Elementary nucleation theory suffices for
this purpose.  Equilibrium crystal surfaces undergo well defined
roughening transitions from macroscopic flat to macroscopic rough.  On
the other hand, growing surfaces are theoretically ``always
rough''~\cite{Nozieres,Villain}.  This seems at odds with practical
reality, where surfaces appear to grow quite differently below and
above the equilibrium roughening temperature $T_{\rm R}$.  Above
$T_{\rm R}$ they are rough (dynamic roughness) while below $T_{\rm R}$
they seem flat (layer-by-layer step-flow growth).  Above $T_{\rm R}$
the growth velocity $v_{\rm g}$ is proportional to the oversaturation
$v_{\rm g}\sim\Delta \mu$, while below $T_{\rm R}$ it is inversely
proportional to a nucleation time scale $v_{\rm g}\sim \tau^{-1}$ with
$\tau^{-1} \sim \exp\left({-a \eta^2/ (\Delta \mu
    k_BT)}\right)$~\cite{Nozieres}.  $\eta$ is the equilibrium step
free energy.  As a result, crystal growth shapes have sharp angles, in
which many facets, including all that are above their $T_{\rm R}$, are
missing.  This apparent difference in growth mechanism is one of the
most useful experimental tools to locate equilibrium roughening
transitions in crystal facets.

The origin of the exponential factor in $\tau$ is the existence of a
nucleation barrier for creating a terrace of height $h\to h+1$ below
$T_{\rm R}$ .  The edge (step) free energy loss term (proportional to
$\eta$ times the circumference) competes with the surface energy gain
term (proportional to $\Delta \mu$ times the terrace area). The
nucleation barrier vanishes when the step free energy $\eta$ vanishes,
i.e., at $T_{\rm R}$.  After a new terrace larger than the nucleation
thresholds is nucleated with an exponential small probability, it
spreads out fast by particle adhesion at its edge into a macroscopic
domain, until it merges with other spreading terraces that have
nucleated in the mean time, and thus complete the new surface layer.
However, new terraces are nucleated on top of
spreading terraces as well.  This nesting effect, together with the spatial
fluctuations of nucleation events leads to a loss of a well defined
(length scale free) global reference surface level.  This means that
although at small enough length scales the surface looks flat and
seems to grow layer-by-layer, at large length scales it is rough.

There is no phase transition between the layer-by-layer and rough
growth regimes, only a characteristic crossover length scale.  The
latter is of order $l_{\rm R}= v_{\rm s}\tau$, with $v_{\rm s}$ the
step velocity (determined by the particle deposition rate at the step
edge) and $\tau$ the above time scale at which terrace nuclei are
being created.  Surface flatness cannot be maintained during growth
over large length scales, but at small oversaturations ($\Delta \mu$)
and sufficiently below $T_{\rm R}$ (large step free energies $\eta$)
the growing surface can appear to be flat for all practical purposes,
over any typical experimental length scale.

The same type of issues arise in our study concerning the
compatibility of surface reconstruction order with growth dynamics.
First we address whether surface reconstruction order can persist
during growth at macroscopic length scales (the thermodynamic limit);
and, if not, whether it might still exist in a practical sense within
a characteristic length scale, $l_{\rm rec}$ below the equilibrium
reconstruction temperature $T_{\rm rec}$.

\section{Reconstructed Rough Equilibrium Phases}\label{recrough}

Surface reconstruction is conventionally associated with flat
interfaces.  However, surface roughness not necessarily destroys the
reconstruction order.  A rough but still reconstructed surface is in a
so-called reconstructed rough (RR) phase.  The equilibrium versions of
RR phases were studied theoretically some years ago in the context of
the competition between surface roughening and reconstruction in MR
reconstructed FCC (110) facets~\cite{VV,MdN-Pt}.  The topological
details of those FCC surface prevent the existence of RR phases,
implying that in Pt(110) the surface roughness and deconstructs
simultaneously~\cite{MdN-Pt} as observed experimentally in
Pt(110)~\cite{RVK}.  This implies immediately that during growth
reconstruction order is limited to the roughness crossover length
scale, $l_{\rm rec}\leq l_{\rm R}$.  The same theoretical studies 
also identified other surface geometries where RR phases do exist.
For those $l_{\rm rec}$ is not limited by $l_{\rm R}$.  In this
section we review the basic properties of RR phases, using as examples
checkerboard and MR type reconstructed simple cubic stackings.

\begin{figure}
  \centerline {\epsfxsize=5 cm \epsfbox{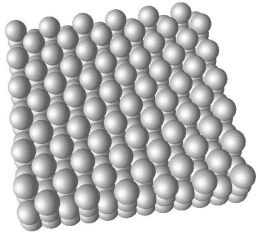}} \vskip 10 true pt
\caption{
  Checker board type misplacement surface reconstruction}
\label{chkbrd}
\end{figure}

To avoid confusion, it is useful to distinguish between misplacement
and displacement type reconstruction~\cite{MdN-King}.  In
misplacement reconstructions, particles have moved to different
solid-on-solid type stacking positions, or are removed altogether,
compared to the unreconstructed flat surface structure.  The checker
board reconstruction in Fig.\ref{chkbrd} and also the more realistic
MR type reconstructions are examples of this.  The average surface
height has changed by half a unit, $h\to h-\frac{1}{2}$.  In
displacement reconstructions the atomic stacking does not change.
Instead, the atoms are merely elastically distorted at the surface
with a commensurate or incommensurate period compared to the bulk.
Misplacement type reconstructions are more likely to disorder at
temperatures near $T_{\rm R}$ than displacement type reconstructions.
For clarity we focuses here on misplacement reconstructions.

\begin{figure}
  \centerline {\epsfxsize=5 cm \epsfbox{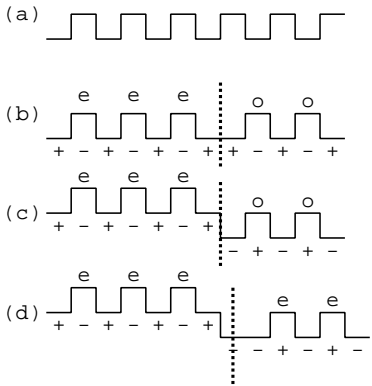}} \vskip 10 true pt
\caption{
  (a) A perfect reconstructed surface. (b) A domain wall; both order
  parameters change sign. (c) A step where only the even-odd row order
  changes sign. (d) A step where only the parity order changes sign.
  }
\label{scmr}
\end{figure}

The definition of the reconstruction order parameter is at the core of
RR phases ~\cite{MdN-Pt,MdN-King}.  In checker board and MR reconstructed SC
(110) facets , the reconstruction order can be formulated in two
distinct ways.  One formulation keeps track of whether the black or
white fields (even or odd rows) are on top.  The other measures it in
terms of anti-ferromagnetic order in the parity type Ising variables
$S_r=\exp(i \pi h_r)$, with $h_r=0,\pm1,\pm2,\cdots$ the surface
height at site $r$, see Fig.\ref{scmr}.  These two formulations might
seem equivalent in flat surfaces, but they are not in the presence of
roughness.

The compatibility of surface reconstruction with surface roughness
depends on topological properties of step and domain wall excitations;
on how they affect the two versions of the order parameter.
Fig.\ref{scmr} shows in cartoon style a cross section of the
reconstructed surface, and also domain wall and step excitations.  The
domain wall in (b) does not change the surface height.  Notice that
both order parameters change sign.  Across the step in (c), from left
to right, the even-odd order parameter changes sign, but the parity
order is unaffected.  At the step in (d) the opposite happens.

These two types of steps are the only topologically distinct ones that
are possible; (c) couples only to the even-odd row type order
parameter and (d) only to to the parity version.  

It is possible to construct many more step and domain wall structures
that look locally different from the ones in the figure, but those
induce the same change in height and/or reconstruction order(s) and
therefore are from a topological point of view identical to the ones
in the figure.  Notice also that the excitations in (b)-(d) are
related to each other in the sense that any of the three can be
interpreted as a bound state the two others. Elastic surface
deformations in the actual atomic positions near the surface and
additional ones near the steps and domain walls, influence the local
internal structure of steps and domain walls, but do not affect those
topological features, and therefore need not explicitly be represented
in the following discussion.  (They certainly renormalize the step and
domain wall energies and the interactions between such surface
excitations.)

The fate of reconstruction versus roughness depends on the energies of
these steps and domain walls, including the kink energies.  They set
the scale of the meander type entropy and therefore the temperature
dependence of the step free energies.  If the domain wall free energy
vanishes first, the surface remains flat but the reconstruction
vanishes, $T_{\rm rec}<T_{\rm R}$.  In case the free energy of one of
the two types of steps vanishes first, the surface enters a
reconstructed rough phase, $T_{\rm R}<T_{\rm rec}$.  At the roughening
transition one of the two reconstruction order parameters vanishes,
but the other type of order remains.  So there exist two topologically
distinct types of RR phases.  (Notice that only the one with the
parity type order is readily observable by, e.g., conventional X-ray
diffraction.)

In the RSOS model below, the RR phase has parity order, i.e., the step
free energy of the (c) type steps is zero, but walls and (d) type
steps still have non-zero free energy.  We will refer to those
excitations as ``loops of zero's", because in the rough surface they
show up as contours across which the height change is zero, $dh=0$.
The deconstruction transition (inside the rough phase) takes place at
the temperature where the surface tension of the loops vanishes.  In
equilibrium that turns out to be an ordinary Ising transition.  This
conclude our brief review.  For more details we refer to
Refs.~\cite{MdN-Pt} and \cite{MdN-King}.

\section{reconstructed rough growth}\label{recroughgrowth}

Let's focus now on surface growth.  Only in surfaces where equilibrium
RR phases are topologically possible, can the surface reconstruction
length scale $l_{\rm rec}$ exceed the onset of dynamic roughness
length scale $l_{\rm R}$.  Moreover, it is quite possible that the
reconstruction order persists over all length scales ($l_{\rm rec}\to
\infty$), such that a genuine dynamic deconstruction phase transition
takes place in the stationary state of the growing surface, just like
in equilibrium.

For comparison, imagine a two dimensional (2D) lattice with on each
site an height variable and an Ising spin degree of freedom
(representing the reconstruction order). This leads to two coupled
master equations, one for surface growth, e.g., KPZ type dynamics, and
another for the reconstruction order, e.g., Glauber type Ising
dynamics.  In equilibrium surfaces, the coupling between the two
sectors is weak, to the extend that the reconstruction transition in
the Ising sector and the roughening transition in the height variable
sector do not interfere with each other~\cite{MdN-Pt,MdN-RSOS}.  The central
issue is whether and how this coupling changes during growth.
The Ising dynamics itself, is blind to the growth bias.  If the
coupling between the two sectors is remains weak, the Ising spins can
still reach the Gibbs equilibrium state and undergo a conventional
equilibrium reconstruction transition.

Coupled master equations of this type have been studied recently in
the context of specific 1D growth models.  Those display strong
coupling between the Ising and roughness degrees of freedom, such as
growth being pinned down by Ising domain
walls~\cite{Drossel,Kotrla,Noh}.  Pinning favors to spontaneous
facetting. In our 2D model, we observe different effects, besides the
obvious fact that in 1D equilibrium reconstruction order can not
exist.

\section{ Restricted solid-on-solid Model}\label{RSOSmodel}

The 2D restricted solid on solid (RSOS) is one of the work horses of
surface physics research. Integer valued height variables $h_r=0,\pm
1,\pm2,\cdots$ are assigned to a square lattice and nearest neighbor
heights are restricted to differ by at most one unit, $dh=0,\pm1$. The
energy
\begin{equation}
E =  \frac{1}{2} K \sum_{\left<r, r^{\prime}\right>} (h_r-h_{r^{\prime}})^2
\end{equation}
depends only on nearest neighbor interactions.  We use dimensionless
units, $K=J/k_{\rm B}T$.  The $K>0$ side of the phase diagram contains
a conventional equilibrium surface roughening
transition~\cite{MdN-RSOS}.  Moreover, the non-equilibrium version has
been studied extensively for $K>0$ as well, because it is a natural
lattice realization of KPZ growth ~\cite{KPZ,HHZh,RSOS,KrSp,KANK,KK,CCS-MdN}.

For $K<0$, the model contains one of the simplest examples of an
equilibrium RR phase~\cite{MdN-RSOS}, and is probably the most compact
formulation of the coupling between Ising and surface degrees of
freedom.  The $dh=\pm1$ steps are more favorable than flat $dh=0$
segments.  At zero temperature, $K\to-\infty$, the $dh=0$ states are
frozen out, and the model reduces to the so-called body centered solid
on solid (BCSOS) model, but in this version it lacks step energies,
which means that the surface is rough even at zero temperature.  The
surface is rough, but since nearest neighbor heights must differ by
one, all heights on one sublattice are even and odd at the other, or
the other way around.  This two-fold degeneracy represents the checker
board type RR order.  The staggered magnetization, defined in terms of
the parity spin type variables $S_i = \exp(i\pi h_r)$, is non-zero.

The $dh=0$ excitations that appear at $T>0$ form closed loops and
behave like Ising type domain walls. The reconstruction order changes
sign across such loops.  Their sizes diverge at the equilibrium
deconstruction transition $K_{\rm c}=-0.9630$~\cite{MdN-RSOS}.
(Determined by transfer matrix finite size scaling techniques).  The
Ising and roughness variables couple only weakly.  Numerically, all
reconstruction aspects of the transition follow conventional Ising
critical exponents.  Moreover, the thermodynamic singularities in the
Ising sector affect only the temperature dependence of the surface
roughness parameter $K_{\rm G}$, defined in terms of the height-height
correlator,
\begin{equation}
\langle (h_{r+r_0}-h_{r_0})^2 \rangle \simeq (\pi K_{\rm G})^{-1} \ln(r).
\end{equation}
The continuum limit analysis confirms these numerical results.  The
point in the generalized phase diagram where the Gaussian (height) and
Ising degrees of freedom decouple is a stable renormalization type
fixed point~\cite{MdN-King}.

We study this same RSOS model in the presence of a KPZ type growth
bias.  In the MC simulation, we first select an update column and next
whether a particle deposition or evaporation event will be attempted.
The move is rejected if it would result in a violation of the RSOS
condition, $dh=0,\pm1$.  If allowed, it will take place with
probability $P =\min(p, p e^{-\Delta E_j})$ in case of deposition, and
with probability $P =\min(q, q e^{-\Delta E_j})$ for evaporation.
Without loss of generality we can choose $p+q=1$.  At infinite
temperature ($K=0$) and deposition only ($q=0$) the model reduces to
the well known Kim-Kosterlitz ~\cite{KK} model for KPZ type growth.

We will present only our MC results far from equilibrium, i.e., at
$q=0$ with deposition only.  We observe no qualitative differences
closer to equilibrium, $0<p<1$, but the interpretation of the data
becomes increasingly obscured (as expected) by (conventional)
crossover scaling from the equilibrium deconstruction phase
transition.

At low temperatures, $K \rightarrow -\infty$, the Metropolis dynamics
slows down considerably. The rejection rate becomes high and the
density of active sites becomes low.  Therefore we employ the
following rejection free algorithm.  During the MC simulation we keep
a list of active sites, i.e., sites where particles can deposit
without violating the RSOS condition.  They are grouped in
$j=1,\cdots,5$ sets, according to the five distinct energy changes
$\Delta E_j$ that can occur during deposition.  First we preselect one
of those 5 sets, with probability $(p_j N_j)/(\sum_j p_j N_j)$, where
$p_j =\min(1,e^{-\Delta E_j})$ and $N_j$ is the number of sites of
type $j$.  Next, a particle is randomly deposited at one of the sites
in that specific set $j$.  Rejection free procedures like this upset
the flow of time.  To restore proper time, we increase the MC time
during each update step by $1/p\times 1/N_j$.  We checked explicitly
that this reproduces the correct value for the KPZ dynamic exponent
$z=8/5$~\cite{KK,CCS-MdN,Lassig} at $K\simeq 0$; we find $z\simeq
1.6\pm 0.1$.

The above algorithm resolves the slowing down problem in the actual MC
simulation, but does not address its origin.  In the limit $K
\rightarrow -\infty$ the RSOS model reduces to the BCSOS model, with
$dh=\pm1$ at all bonds.  The $dh=0$ loops are frozen out completely.
In BCSOS type KPZ growth dynamics, 2 particles are deposited at ones
in the form of vertically oriented bricks, otherwise a ``forbidden''
configuration with $dh=0$ would arise.  In the $K<0$ RSOS model at
very low temperatures the same event is achieved as a 2-step
two-particle process, by the deposition of a second particle at the
same site soon after the first one.  The probability for deposition of
the first particle is equal to $p= L^{-2}\exp(2K)$.  The second
particle deposition on top of it happens with probability $p=L^{-2}$).
This implies that the time clock in the RSOS model runs slower by a
factor $r=\exp(2K) ( 1 + 4 \exp(K) + ...)$.

A final remark about surface roughness.  In normal surfaces, the
equilibrium roughness increases with temperature; due to the fact that
meander type entropy renormalizes the step energy into a reduced step
free energy~\cite{MdN-King}.  In our model, surface roughness evolves
the opposite way; it decreases with temperature.  The surface is less
rough at infinite temperature $K=0$ than in the zero temperature limit
$K\to-\infty$.  A high temperature RSOS surface, with $dh=0,\pm1$ is
obviously less rough than a BCSOS surface, with only $dh=\pm1$.
Recall that this BCSOS model lacks step energies, such that it is just
as rough at $T=0$ as at $T\to\infty$.  From the BCSOS perspective the
thermally excited $dh=0$ loops stiffen the surface, and give rise to
an inverted roughness versus temperature profile. On the one hand,
this is an interesting phenomena in its own right.  Moreover we could
fine tune it by introducing next-nearest neighbor interactions, since
they represent BCSOS type step energies.  On the other hand, this
effect is unlikely to affect the central question we want to address
(how do roughness and reconstruction degrees couple during growth) and
therefore we choose not do so in this study.

\section{ reconstruction during growth}\label{MCdata}

\begin{figure}
  \centerline{\epsfxsize=8cm \epsfbox{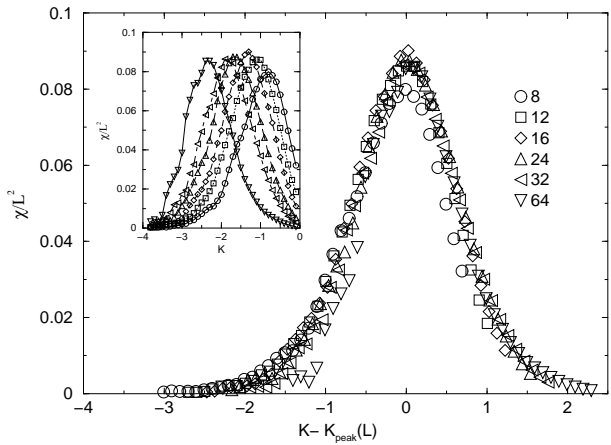}} \vskip 10 true pt
\caption{
  Reconstruction order susceptibility $\chi$ as function of
  temperature at system sizes $L=8$-$64$.  The data collapses onto a
  single curve by the shift $K^\prime = K-K_{\rm{peak}}(L)$, with
  $K_{\rm{peak}}(L)=-0.77\ln(L/2.2)$.}
\label{susClps}
\end{figure}

We search for reconstruction order as function of temperature, for
$-\infty<K<0$.  The susceptibility type parameter~\cite{MC-suscep},
\begin{equation}
\chi = L^2(\langle m^2 \rangle - \langle | m | \rangle^2)
\end{equation}
of the reconstruction order parameter,
\begin{equation}
m=\left< (-1)^{x+y} ~e^{i \pi h(x,y)} \right>
\end{equation}
is shown in Fig.\ref{susClps} for the stationary state of the growing
surface, as function of $K$ for different system sizes $L^2$.  The
sharp maxima seem to confirm the existence of a dynamic surface
reconstruction transition into a RR phase.  However, several features
are very different from equilibrium. The peak height diverges as
$\xi\sim L^2$. i.e., stronger than at the equilibrium transition point
where it scales as $\chi\sim L^{\gamma/\nu}$.  This could be a signal
of a first order phase transition.  However, the peak position does
not converge to a specific critical point $K_c$. Instead it keeps
shifting with lattice size. It scales logarithmically, as $K_{\rm
  peak}(L)\simeq-A\ln(L/L_0)$ with $A=0.77\pm 0.05$ and $L_0= 2.2\pm
0.2$.

\begin{figure}
  \centerline {\epsfxsize=8 cm \epsfbox{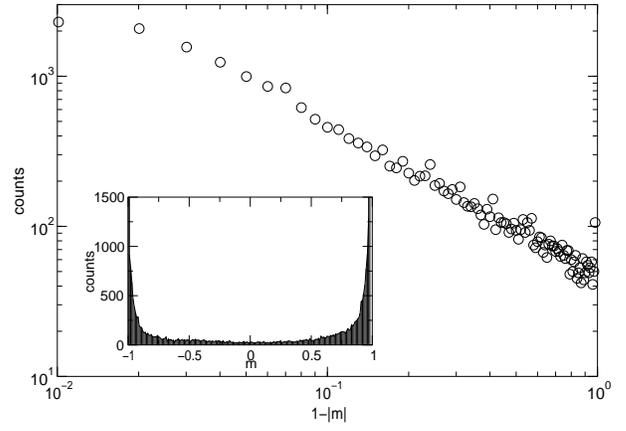}} \vskip 10 true pt
\caption{
  Histogram (insert) of the reconstruction order parameter, $m$, at
  $L=32$ and $K=-3.2$ from $2^{18}$ data points using $\Delta M=0.01$
  as bin width.  The tails about the peaks at $m=\pm 1$ scale as power
  laws (main frame) with exponent $-0.9\pm0.1$.}
\label{mhstgm}
\end{figure}

Next, we monitor in detail the reconstruction order parameter $m$ near
and below the equilibrium $K_c$ as function of time.  It behaves
similar as in conventional spontaneously ordered phases, but
flip-flops more frequently than justifiable from finite size effects
alone. Moreover, the fluctuations in $m$ within each phase are too
strong.  Fig.\ref{mhstgm} quantifies this in terms of a histogram of
the number of times a specific value of $m$ appears in a typical time
series.  The distribution has two distinct peaks, suggesting the
presence of spontaneously broken reconstruction order, but the tails
have a power law shape instead of the exponential form mandatory for a
spontaneous broken symmetry.

Power laws are the hallmark of critical fluctuations.  So, quite
surprisingly, it appears as if the RR order is critical at low
temperatures for all $K<K_{\rm peak}$. Instead of an isolated critical
point, we seem to be dealing with a critical phase.

\section{ Loops trapped on ridge lines }\label{trappedloops}

\begin{figure}
  \centerline{\epsfxsize = 8 cm \epsfbox{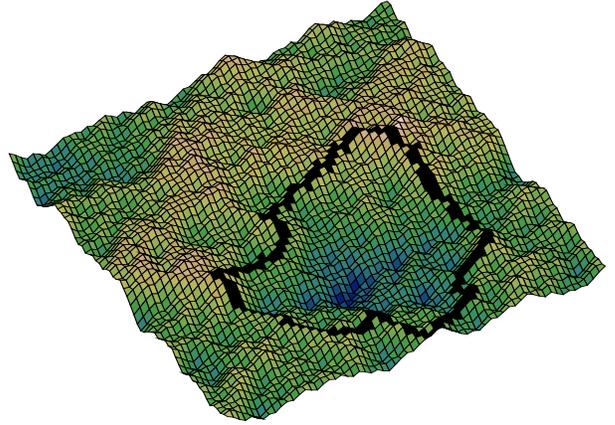}} \vskip 10
  true pt
\caption{ A typical low temperature configuration of the
  growing surface with one large loop trapped at a ridge-line.  }
\label{trappedloop}
\end{figure}

The surprising critical fluctuations in the reconstruction order
parameter can be traced to the following loop dynamics.  Consider a
typical configuration at very low temperatures.  Fig.\ref{trappedloop}
shows an example~\cite{MdN-Korea}.  The surface is in an almost pure
BCSOS type dynamic rough stationary state (with $dh\pm 1$), and
contains only a few $dh=0$ loops separating surface areas of opposite
checker board type RR order.

The typical life cycle of such a loop runs as follows. It is nucleated
in a valley bottom. Next it runs up hill, grows in diameter,
encompassing the entire valley, until it becomes trapped on a ridge
line.  There it lingers until another loop annihilates it, or when the
KPZ surface fluctuations to which it is slaved shrink it back to zero.

\begin{figure}
  \centerline{\epsfxsize = 8 cm \epsfbox{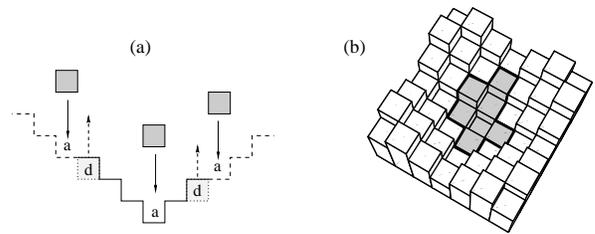}} \vskip 10 true pt
\caption{ 
  (a) One dimensional cross section of the surface near a valley with
  two loop segments.  On the slope, {\it a} ({\it d\,}) are the only
  active adsorption (desorption) sites.  The domain walls always move
  upwards during adsorption. (b) A loop of size of $l_c$ nucleated at
  the bottom of a local valley. Gray and white sites have different
  surface reconstruction parity order.}
\label{1D_dyn}
\end{figure}

Fig.\ref{1D_dyn}(a) represents a cross-section of the 2D rough surface
near a valley.  It shows a domain of opposite reconstruction inside an
otherwise perfectly reconstructed rough configuration.  The two flat
segments are the locations where the domain wall loop intersects the
cross section.  In equilibrium, the loop fluctuates with equal
probability up and down the slope because depositions and evaporations
are equally likely.  A growth bias breaks this symmetry, the loops
move more likely upwards than downwards, see Fig.\ref{1D_dyn}(a).  This
upward drift is the driving force responsible for the trapping of
loops at ridge lines, and thus creates a strong coupling between the
roughness and reconstruction degrees of freedom, unlike equilibrium
where they effectively decouple.

A few comments on the topology of ridge lines in rough surfaces might
be useful. Imagine a rolling ball in this landscape, like in the well
known analogy with renormalization flow in statistical physics.
Presume strong friction such that the velocity is proportional to the
force, i.e., the gradient of the slope, at all times.  The hill tops
are the completely unstable ``fixed points''.  The valleys are the
attractors.  The ridge lines form the water sheds between valleys.
Every ridge line runs from an hill top to a saddle point.  From each
hill top an arbitrary number of ridge lines can emerge, but only two
ridge lines can end at each saddle point (at opposite sides of the
single direction in which the saddle point attracts).  So the ridge
lines form a network, and since none of them can not stop in midair it is a
closed network.  The KPZ rough surface is scale invariant, which means
that this ridge line network has fractal properties.

Ignore for the time being the scale invariant aspects of the network.
Imagine a landscape consisting of deep smooth valleys surrounded by
ridge lines; unlike the real rough surface where every deep valley
consists of collections of sub valleys.  The life cycle of a
macroscopic loop in this surface starts with the nucleation of a new
seedling-loop at the floor of the valley and its rise along the
slopes, during which it grows into a macroscopic object.  The only
loops of interest are those nucleated at the valley bottom and then
run up-hill encompassing the entire valley.  Only those loops are
topologically trapped and stable.  Loops nucleated on the slopes
annihilate by stochastic fluctuations before becoming macroscopically
large. The same is true for loops nucleated out of the valley bottom
but running up-hill on one slope segment only.

The rise of a seedling loop out of the valley bottom into a
macroscopic object, is a very fast process.  Almost no MC moves that
make the loop grow and rise are rejected; energy barriers are rare,
because the length of the loop (its energy) increases uniformly.
Compared to this, the nucleation frequency in the valley bottom is
very small.  This means that the time scale at which a macroscopic
loop emerges out of the valley is limited by the nucleation time scale
$\tau_{\rm n}$ and independent of the valley size.

To measure $\tau_{\rm n}$ we prepared a surface in the BCSOS KPZ
stationary state and measure (at a very low temperature, $K \ll K_{\rm
  c}$) the intervals between macroscopic loop events.  Numerically we
find $\tau_n\sim\exp(- \alpha K)$ (measured in BCSOS time units) with
$\alpha=3.0\pm 0.1$.

This agrees qualitatively with the following estimate.  The deposition
of the first particle in the valley bottom occurs with probability
$p=L^{-2}e^{2K}$.  This creates a fledgling loop, but one that is
indistinguishable from the intermediate state in an elementary BCSOS
type growth event (where a second particle is dropped on top of it
with probability $p=L^{-2}$).  The loop grows when the next particle
is dropped not on top but next to the previous one.  That happens with
probability $p=L^{-2}e^{K}$.  The nucleation threshold diameter
$l_{\rm c}$ is reached when the loop growth and BCSOS growth become
distinguishable, i.e., when the annihilation of a loop requires the
creation of a new well distinguishable loop inside it.  That happens
at about $l_c^2 \simeq 7$, see Fig.\ref{1D_dyn}b.  The time scale at
which that stage is reached is approximately $t\simeq L^{-2}e^{-4K}$
(in BCSOS time units), which is of the same order of magnitude as the
above numerical nucleation time scale.

The loop rises out of the valley until it becomes trapped on the ridge
line that borders this valley to adjacent ones.  From
there on the loop is slaved to the growth fluctuations of the surface.
Valleys grow and shrink (without bias), open-up, fill-up and merge.
The loop has to follow this dance of the ridge line until a new loop
nucleates out of the valley and annihilates it, or when the encircled
terrain happens to shrink to zero (fills-up) by surface growth
fluctuations.

\begin{figure}
  \centerline{\epsfxsize = 8 cm \epsfbox{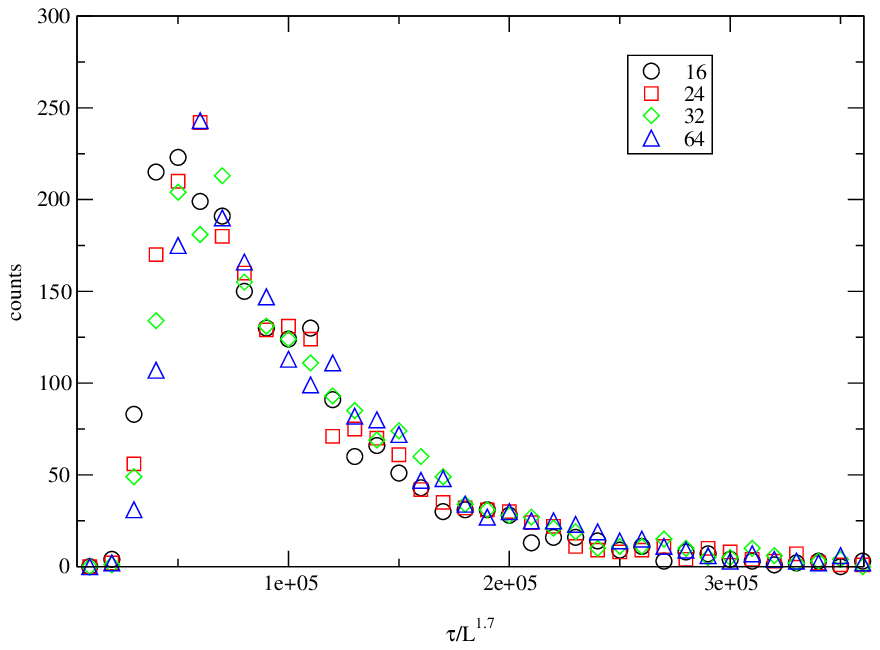}} \vskip 10 true
  pt
\caption{
  Histogram of the decay time of a trapped loop at $K=-6.0$.  The data
  collapses by rescaling time by a factor $L^{1.7}$.} .
\label{timeHis}
\end{figure}

We expect that the life time $\tau_z(L)$ of a ridge line of size $L$
in a growing surface, scales as a power law, $\tau_z\sim L^z$, with
$z$ the dynamic exponent of the surface roughness degrees of freedom
(KPZ like in our model).  To test this, we measure the decay times of
large macroscopic defect loops (of about half the lattice
size) as function of $L$, at low temperatures $K\ll K_c$.  The data in
Fig.\ref{timeHis} collapse indeed onto one universal curve after a
rescaling of time by $\tau_z \approx L^z$.  The collapse fits best at
$z=1.7\pm 0.1$ (in BCSOS time units), which is consistent with the
known KPZ dynamic exponent $z=8/5$ \cite{KK,CCS-MdN,Lassig}.

The ridge line fluctuations are responsible for the power law tails in
the time distribution of RR order, Fig.\ref{mhstgm}.  Those critical
fluctuations only show up below a characteristic length scale $l_{\rm
  rec}$, where the nucleation time scale $\tau_{\rm n} \sim
\exp(-\alpha K)$ is larger than the surface growth time scale, $\tau_z
\sim L^z$.  A simple estimate for $l_{\rm rec}$ follows from equating
the two time scales, $l_{\rm rec}\sim \exp(\frac{\alpha}{z} K)$.

The peaks in the susceptibility, in Fig.\ref{susClps}, reflect this
crossover length $l_{\rm rec}$.  Recall that the peak shifts
logarithmically. By setting $\tau_n=\tau_z$ we obtain the same
logarithmic behavior, $K_{\rm c}=-\frac{z}{\alpha}\ln(L/L_0)$.  The
prefactor is too small by about 30\%, but this is not a surprise
because the estimate is rather simple minded.  It ignores for example
the self similarity of the rough surface.  Consider a sub valley
adjacent to an already trapped loop.  Suppose a new loop nucleates out
of this subvalley.  The loop segments annihilate each other in pairs.
The net effect of this nucelation event is therefore that the trapped
loop jumps across the sub valley.  It now follows the complementary
segment of the ridge line that encircles the sub valley.  Such events
renormalize $\tau_z$, in particular near $K_c$.

\section{ Conclusions }\label{conclude}

In this paper we study the compatibility of surface reconstruction and
surface roughness during growth.  There are several possibilities.

In surfaces where reconstructed rough (RR) phases are topologically
forbidden, like missing row reconstructed FCC(110) facets, the
reconstruction order can not exist on a global scale in the stationary
growing state.  It can appear only locally within the crossover
roughness length scale $l_{\rm R}$ within which the surface grows in a
layer-by-layer fashion, i.e., $l_{\rm rec}\leq l_{\rm R}$.

The reconstruction length scale $l_{\rm rec}$ can exceed $l_{\rm R}$,
only in surfaces where equilibrium reconstructed rough phases are
topologically possible, and those surfaces could in principle even
display genuine deconstruction type phase transitions in the
stationary growing state.

We address this issue in the context of KPZ type dynamics, in the RSOS
model with negative coupling constant $K<0$, which in equilibrium has
a checker board type RR phase and a true deconstruction phase
transition inside the rough phase.  We find that the stationary
growing rough state lacks true macroscopic RR order; $l_{\rm rec}$
remains finite.  Moreover, we identify the mechanism that sets the
temperature dependence of $l_{\rm rec}$.

The fundamental features are an upward drift of the reconstruction
domain wall loops and their trapping at the ridge lines of the
surface.  There, the loops are slaved to fluctuations of the surface growth
dynamics.  $l_{\rm rec}$ is set by the competition between two life
times: the nucleation time scale of a new loop out of the valleys
(annihilating existing trapped loops) and the time scale
$\tau_{KPZ}\sim L^z$ at which a ridge line of radius $L$ vanishes due
to surface growth fluctuations.

At length scales smaller than $l_{\rm rec}\sim \exp(\frac{\alpha}{z}
K)$, the surface appears as reconstructed rough, and the life time of the
loops is determined by the KPZ growth dynamic fluctuations.  The
latter follow power laws. This manifests itself in critical
fluctuations in the reconstruction order at length scales smaller than
$l_{\rm rec}$.  In x-ray diffraction from such a growing interface,
one would observe not only power law shaped peaks associated with the
surface roughness, but also at temperatures where $l_{\rm rec}$ is
larger than the coherence length of the surface, power law shaped
reconstruction diffraction peaks.

At length scales larger than $l_{\rm rec}$, the surface appears as
unreconstructed rough. Loops of that size die by nucleation of new
loops instead of KPZ surface fluctuations, and they are not trapped
anymore, because loop segments can hop across sub valleys of size $l >
l_{\rm rec}$ by means of nucleation of new loops in sub valleys.

In our study we chose to focus on KPZ type surface growth dynamics,
but we have good reasons to expect that the trapping of domain walls
on ridge lines is a common phenomenon.  In general, the quasi-critical
fluctuations will reflect the dynamic exponent of whatever growth
dynamics is applicable.  In recent studies of 1D models with KPZ and
Ising type coupled degrees of freedom the Ising defects become trapped
in valleys and canyons and thus pin-down the
growth~\cite{Drossel,Kotrla}.  We expect that a tendency towards
facetting instead of ridge line trapping can also be realized in our
2D model by varying the local growth rates.  This research is
supported by the National Science Foundation under grant DMR-9985806.

\pagebreak

\end{multicols}

\begin{references}
\bibitem{vBN} H.~van Beijeren and I.~Nolden, in {\it Structures} {\it
    and} {\it Dynamics} {\it of} {\it Surfaces}, W. Schommers and
  P.~von Blanckenhagen, eds. (Springer, Berlin, 1987).
  
\bibitem{Wortis} M.~Wortis, in {\it Chemistry} {\it and} {\it Physics}
  {\it of} {\it Solid} {\it Surfaces} {\it VII}, R.~Vanselow and
  R.F.~Hove, eds. (Springer, Berlin 1988).
  
\bibitem{Dash} J.G.~Dash, Contemp.~Phys. {\bf 30}, 89 (1989).
  
\bibitem{Dietrich} S.~Dietrich in {\it Phase Transitions and Critical
    Phenomena}, Vol.12, eds C.~Domb and J.~Lebowitz, (Academic Press,
  London, 1988).
  
\bibitem{King} {\em The Chemical Physics of Solid Surfaces and
    Heterogeneous Catalysis}, Vol. 7 edited by D.~King (Elsevier,
  Amsterdam, 1994).
  
\bibitem{VV} I.~Vilfan and J.~Villain, Surf.~Sci. {\bf 199}, L165
  (1988); Phys.~Rev.~Lett. {\bf 65}, 1830 (1990); Surf.~Sci. {\bf
    257}, 368 (1991).
  
\bibitem{MdN-Pt} M.~den Nijs, Phys.~Rev.~Lett. {\bf 66}, 907 (1991),
  and Phys.~Rev.~B {\bf 46}, 10386 (1992).
  
\bibitem{MdN-PR1} K.~Rommelse and M.~den Nijs, Phys.~Rev.~Lett. {\bf
    59}, 2578 (1987); Phys.~Rev. {\bf B 40}, 4709 (1989).
  
\bibitem{MdN-PR2} M.~den Nijs, Phys.~Rev.~Lett. {\bf 64}, 435 (1990).
  
\bibitem{MdN-King} M.~den Nijs, chapter 4 in the ref.[\cite{King}].
  
\bibitem{RVK} I.~K. Robinson, E. Vlieg, and K. Kern, Phys.~Rev.~Lett.
  {\bf 63}, 2578 (1989).
  
\bibitem{MdN-Si} Marcel den Nijs, J.~Phys.~A.{\bf 30}, 397-404 (1997).
  
\bibitem{Nozieres} P.~Nozi\`{e}res and F.~Gallet, J.~Phys.(Paris) {\bf
    48}, 353(1987).
  
\bibitem{Villain} A.~Pimpinelli and J.~Villain, {\it Physics of Crystal
    Growth} (Cambridge University Press, 1997).
  
\bibitem{KPZ} M.~Kardar, G.~Parisi, and Y.C.~Zhang, Phys.~Rev.~Lett.
  {\bf 56}, 889 (1986);
  
\bibitem{HHZh} T.~Halpin-Healy and Y.C.~Zhang, Phys.~Rep. {\bf 254},
  215 (1995).

\bibitem{RSOS} J.G.~Amar and F.~Family, Phys.~Rev.~Lett. {\bf 64}, 543
  (1990), {\it ibid.} {\bf 64}, 2334 (1990);
  
\bibitem{KrSp} J.~Krug and H.~Spohn, Phys.~Rev.~Lett. {\bf 64}, 2332 (1990).
  
\bibitem{KANK} J.~Kim, T.~Ala-Nissila and J.M.~Kosterlitz,
  Phys.~Rev.~Lett. {\bf 64}, 2333 (1990).

\bibitem{KK} J.M.~Kim and J.M~. Kosterlitz, Phys.~Rev.~Lett. {\bf 62},
  2289 (1989).

\bibitem{CCS-MdN} C.S.~Chin and M.~den Nijs, Phys.~Rev.~E.~{\bf 59} ,
  2633-2641 (1999).

\bibitem{Lassig} M.~L\"assig, Phys.~Rev.~Lett. {\bf80}, 2366 (1998).  

\bibitem{MdN-RSOS} M.~den Nijs, J.~Phys.~A {\bf 18}, L549 (1985).  

\bibitem{Drossel} B.~Drossel, and M.~Kardar, Phys.~Rev.~Lett. {\bf
    85}, 614 (2000).
  
\bibitem{Kotrla} M.~Kotrla, and M.~Predota, Europhys.~Lett. {\bf 39},
  251 (1997); M.~Kotrla, F.~Slanina and M.~Predota, Phys.~Rev.~B {\bf
    58}, 10003 (1998).
  
\bibitem{Noh} J.D.~Noh, H.~Park, and M.~den Nijs, Phys.~Rev.~Lett., in
  press.
  
\bibitem{MC-suscep} K.~Binder, and D.~W.~Heermann, {\em Monte Carlo
    Simulation in Statistical Physics} (Springer-Verlag, Heidelberg,
  1997).

\bibitem{MdN-Korea} A color version of this figure, can be found on
  our WEB page, and also in, Marcel den Nijs, Bulletin of the APCTP
  (Korea), Spring 2001 (in press).

\end{references}
\end{document}